\documentclass[a4paper,fleqn]{cas-dc}

\usepackage{amsmath,amssymb,amsfonts}
\usepackage{algorithmic}
\usepackage{graphicx}
\usepackage[square,numbers]{natbib}
\usepackage{textcomp}
\usepackage{xcolor}
\usepackage{hyperref}
\usepackage{subcaption}
\usepackage{tabularx}
\usepackage{fancyhdr}
\usepackage{lastpage}

\newcommand{\method}{\mbox{\textsc{Sen4x}}}
\newcommand{\methodlate}{\mbox{\textsc{Sen4x}[\textit{late}]}}

\def\BibTeX{{\rm B\kern-.05em{\sc i\kern-.025em b}\kern-.08em
    T\kern-.1667em\lower.7ex\hbox{E}\kern-.125emX}}
\begin{document}

\shorttitle{Beyond Pretty Pictures}
\shortauthors{A. Retnanto et~al.}

\title [mode = title]{Beyond Pretty Pictures: Combined Single- and Multi-Image Super-resolution for Sentinel-2 Images}

\author[1]{Aditya Retnanto}[orcid=0009-0009-2833-5949]
\ead{aretnanto.consultant@adb.org}

\author[1]{Son Le}[orcid=0009-0005-1101-6280]
\ead{shle.consultant@adb.org}

\author[1]{Sebastian Mueller}[orcid=0009-0009-2493-9702]
\ead{smueller.consultant@adb.org}

\author[2]{Armin Leitner}
\ead{leitner@geoville.com}

\author[2]{Michael Riffler}
\ead{riffler@geoville.com}

\author[3]{Konrad Schindler}[orcid=0000-0002-3172-9246]
\ead{schindler@ethz.ch}

\author[1]{Yohan Iddawela}[orcid=0009-0008-7132-2126]
\ead{yiddawela@adb.org}

\address[1]{Asian Development Bank, Philippines}
\address[2]{GeoVille Information Systems and Data Processing GmbH, A-6020 Innsbruck, Austria}
\address[3]{ETH Zürich, Switzerland}


\maketitle

\setlength{\footskip}{30pt}  

\makeatletter
\fancypagestyle{myfooterstyle}{
  \fancyhf{}
  \fancyfoot[L]{A. Retnanto et~al.}
  \fancyfoot[R]{Page \thepage\ of \pageref{LastPage}}
  \renewcommand{\headrulewidth}{0pt}
  \renewcommand{\footrulewidth}{0.2pt}  
}
\makeatother

\pagestyle{myfooterstyle}
\thispagestyle{myfooterstyle}

\begin{abstract}
Super-resolution aims to increase the resolution of satellite images by reconstructing high-frequency details, which go beyond naïve upsampling. This has particular relevance for Earth observation missions like Sentinel-2, which offer frequent, regular coverage at no cost; but at coarse resolution. Its pixel footprint is too large to capture small features like houses, streets, or hedge rows. To address this, we present \method{}, a hybrid super-resolution architecture that combines the advantages of single-image and multi-image techniques. It combines temporal oversampling from repeated Sentinel-2 acquisitions with a learned prior from high-resolution Pl\'eiades Neo data. In doing so, \method{} upgrades Sentinel-2 imagery to 2.5$\,$m ground sampling distance.  We test the super-resolved images on urban land-cover classification in Hanoi, Vietnam. We find that they lead to a significant performance improvement over state-of-the-art super-resolution baselines.
\vspace{1em} 
\end{abstract}

\section{Introduction}
Satellite image super-resolution (SR) seeks to enhance the spatial resolution of satellite imagery by reconstructing high-frequency details. For certain use cases, SR offers a cost-effective alternative to expensive and less regularly captured high-resolution (HR) satellite data. While HR imagery provides unmatched detail, it has two main limitations. First, its high cost is prohibitive for many applications. Second, coverage is uneven: (urban) regions in Europe and North America are revisited frequently, whereas other parts of the world are imaged rarely, if at all, unless specifically tasked at even higher cost. In contrast, low-resolution (LR) images, such as those from Sentinel-2\footnote{Sentinel-2 is often referred to as moderate- or even high-resolution data in the context of satellite imagery, where ground sampling distances can be as large as several kilometers. For our purposes, we refer to Sentinel-2 as the LR image and to PNEO as the HR image for clarity.}, are freely available and offer regular worldwide coverage. This raises a natural question: \emph{To what extent can SR images substitute HR images for downstream analysis?}

There has been some skepticism about the practical value of SR, with concerns that it may only improve the visual appearance without adding meaningful information. Reconstructing unobserved high-frequency detail inevitably runs the risk of introducing artifacts. The likelihood of such artifacts depends on the specific SR approach used. Moreover, it is not always clear whether they affect subsequent analysis tasks. Broadly, SR methods can be grouped into two types:

\begin{itemize}
    \item \textit{Multi-image super-resolution (MISR)} which leverages subtle differences between multiple acquisitions, due to sub-pixel shifts, to reconstruct fine spatial details.
    \item \textit{Single-image super-resolution (SISR)} which relies on patterns learned from large datasets of HR imagery (i.e. learned priors) to infer and reconstruct fine structures from a single image.
\end{itemize}

While most SR methods adopt either MISR or SISR, the two approaches are complementary, and combining them can potentially bring further improvements. We find that integrating them into a single model indeed produces images that are more useful for downstream analysis.

Ultimately, the goal of SR is to obtain better insights than one could derive from LR images alone. Nevertheless, SR methods are typically evaluated in terms of their visual quality, rather than in terms of their suitability for subsequent analytical tasks such as segmentation, retrieval or object detection. This points to a key challenge: balancing perceptual quality with physical realism. SISR models often produce sharp, visually appealing results, but are prone to artifacts and hallucinated structures that do not depict the real situation. MISR models, on the other hand, tend to remain more faithful to the physical signal by relying on multiple samples of it. However, their outputs can suffer from blur due to averaging effects.

To mitigate their respective weaknesses and find a better trade-off, we introduce \method{}, a hybrid SR model that combines the strengths of MISR and SISR. \method{} fuses multi-pass oversampling with learned priors to achieve high-quality reconstructions that are both sharp and physically consistent.

We apply \method{} to enhance Sentinel-2 imagery from 10 to 2.5 meter resolution and evaluate its usefulness for land-cover (LC) classification, a widely used benchmark task in remote sensing. Specifically, we compare classification performance using three input sources: \emph{(i)} super-resolved Sentinel-2 images; \emph{(ii)} HR imagery from Pléiades Neo; and \emph{(iii)} naïvely upsampled Sentinel-2 images.

In this way, the comparison goes beyond somewhat ill=defined "visual quality" and instead evaluates the utility of SR for subsequent information extraction. Our findings indicate that SR significantly enhances semantic segmentation in our test area in Hanoi, Vietnam. Among the methods tested, our proposed \method{} model yields the best results. It boosts the mean intersection-over-union (mIoU) by 2.7 percentage points compared to only SISR, and by 12.9 percentage points compared to only MISR (see Table~\ref{tab:sr_lulc_metrics}).

Furthermore, we find that traditional image quality metrics like Peak Signal-to-Noise Ratio (PSNR) or Structural Similarity Index Measure (SSIM) can be misleading: they are poor proxies for segmentation performance. In other words, \emph{prettier pictures} according to simple metrics of image quality \emph{are not necessarily more useful pictures}. On the contrary, they may give rise to significantly worse segmentations.

At first glance, it might seem unnecessary to perform SR as a separate step, since one could instead train a model to predict high-resolution land-cover maps directly from low-resolution images. However, there are two reasons why this approach is less effective: \emph{(i)} Learning such a model is more difficult, because it lacks the guidance that HR images provide for SR. This additional training signal does not require manual labeling, it is a form of self-supervision. \emph{(ii)} Treating SR as a separate step has practical advantages: the enhanced images can be used across multiple tasks, including manual interpretation but also automated analysis with lightweight models that need not provide the capacity for SR.

In summary, the contributions of this paper are:
\begin{itemize}
    \item \method{}, a hybrid MISR+SISR architecture for Sentinel-2, whose outputs are particularly well-suited for automated downstream analysis.
    \item An experimental evaluation of recent SR models, with a focus on LC classification instead of task-agnostic image quality.
    \item A new benchmark for 4$\times$ SR of Sentinel-2 images in the RGB and NIR bands, applicable for both the SISR and MISR modes (as well as hybrid designs).
\end{itemize}
Code and trained models will be made publicly available at 
\url{https://github.com/ADB-Data-Division/sen4x}.

\section{Related Works}
SR methods for remote sensing imagery have evolved from early hand-crafted sensor fusion to modern learning-based models. A foundational precursor of SR is pansharpening, where HR panchromatic images are fused with lower-resolution multispectral data to enhance spatial detail while preserving spectral information~\cite{wald1997fusion,aiazzi2002context}. In recent years, pansharpening has also been approached with deep learning tools like Convolutional Neural Networks (CNNs)~\cite{yang2017pannet}.
At the same time, learning-based SR methods have gained traction. These methods learn from large datasets how LR and HR image patches relate to each other, instead of relying on fixed rules. In doing so, they learn priors from the data, which help them predict fine details that simple upsampling methods cannot recover.
 Early machine learning approaches to satellite image SR include sparse coding, support vector regression, and exemplar-based mappings \cite{freeman2002example,li2009super,zhang2014example}. Today, the SR landscape is dominated by deep neural networks.
An important distinction is between single-image methods and methods that ingest multiple images of the same scene. Single-image methods rely solely on patterns learned from training data, which are encoded in the network’s weights. In contrast, multi-image methods can take advantage of slight pixel shifts between repeated observations of the same location, which effectively provide additional spatial detail.
By and large, SISR and MISR have been studied separately, with limited attempts to combine them, e.g,~\cite{richard19_3dv}. In this paper, we put forward a hybrid scheme that combines the two concepts.

\subsection{Single-image Super-resolution}
The emergence of deep learning-based SISR began with early adaptations of CNNs~\cite{liebel2016single}. 
A modified version of the Enhanced Deep Residual Network (EDSR) has also been tailored for Sentinel-2, incorporating additional near-infrared (NIR) bands to improve performance across spectral channels \cite{galar2020super}. More recently, Swin2MoSE introduced a SISR method that is based on the Swin Transformer architecture, and is therefore capable of capturing possible long-range dependencies \cite{rossi2024swin2mosenewsingleimage}.

Generative models have become increasingly prominent in satellite SISR. ESRGAN, a widely used GAN-based model originally developed for natural images, has been adapted to Sentinel-2 data, enabling more realistic texture generation in the absence of ground truth \cite{salgueiro2020super}. Recently, denoising diffusion models have emerged as a new frontier for satellite SR, offering stable training and high-quality reconstructions \cite{donike2025trustworthy,xiao2024ediffsr,saharia2021imagesuperresolutioniterativerefinement}.

A special case that slightly blurs the boundary between SISR and MISR is L1BSR. This approach makes use of subtle pixel shifts that occur in the overlapping regions of Sentinel-2’s CMOS detectors, which are individual sensor components that each capture part of the image. These natural overlaps introduce slight variations, which L1BSR uses to perform self-supervised SR and align spectral bands, without needing HR reference images \cite{nguyen2023l1bsr}.

\subsection{Multi-image Super-resolution}
MISR improves resolution by combining information from multiple LR images of the same area, often taken at different times. One of the first deep learning models to do this was HighResNet, which uses a recursive, pairwise fusion strategy to integrate image sequences \cite{deudon2020highres}. Later versions of HighResNet introduced a radiometric consistency loss to ensure that brightness and spectral values remain consistent across time \cite{razzak2023multi}.
More recent approaches combine convolutional layers with attention mechanisms \cite{okabayashi2024LTAE}. For example, the Lightweight Temporal Attention Encoder has been enhanced with a fusion module that helps align LR and HR images taken at different times, correcting for temporal mismatches \cite{garnot2020lightweight}.
Worldstrat ~\cite{worldstrat} offers an extensive and recent benchmark of existing MISR models, and introduces a new dataset of temporally aligned Sentinel-2 sequences paired with HR SPOT-6/7 images.

SATLAS \cite{wolters2023zoomingzoominginadvancing} achieves good SR performance with a straightforward scheme. It adapts a SISR model for multi-image inputs. Rather than using a specialized fusion module, multiple input images are stacked together and fed into an ESRGAN. The model is then trained on a large dataset to make it broadly applicable across different regions and conditions.

\subsection{Cross-sensor Super-resolution}
Many existing SR models for satellite images are trained on synthetic datasets, in which LR images are created by downsampling the reference HR images \cite{Lanaras2018SuperResolution}.
However, synthetic datasets may not accurately represent the spectral distribution of actual LR data \cite{Dong2022RealWorld}. Furthermore, in the MISR case, synthetic datasets assume that all LR images were acquired under the same atmospheric conditions, which is not realistic for satellite imagery. To address these limitations, cross-sensor datasets pair LR and HR data from different sensors, resulting in SR models that are more robust to spectral, geometric and atmospheric variations
\cite{Chen2021RealWorldSISR}. Despite the challenges of handling sensor differences inherent in cross-sensor datasets, SR frameworks that pair HR data with real LR images (rather than with synthetically downsampled ones) have achieved markedly improved results \cite{Qiu2023CrossSensorSR}. 

\subsection{Evaluation Metrics for Super-resolution}
SR images are typically assessed by comparing them to HR ground truth using pixel-wise metrics. The most common of these is \emph{PSNR} (Peak Signal-to-Noise Ratio), which has a long tradition in signal processing. However, being based on the mean squared error (MSE), PSNR favors smooth outputs that lack fine details, which can negatively affect tasks that depend on texture or edge information. To address this limitation, \emph{SSIM} (structural similarity index, \cite{SSIM}) was introduced. Unlike PSNR, SSIM considers structural information and aligns better, with how humans judge visual quality.

More recently, deep learning-based metrics have been developed to assess perceptual similarity. One of the most widely used is \emph{LPIPS} (Learned Perceptual Image Patch Similarity \cite{LPIPS}). This compares feature representations extracted by deep neural networks and has been shown to correlate well with perceived image quality. Similarly, \emph{CLIPScore} \cite{wolters2023zoomingzoominginadvancing} uses vision-language models to evaluate semantic consistency between images.
Meanwhile, \emph{OpenSR-Test} \cite{aybar2024opensr} attempts to standardize the evaluation of satellite image SR with dedicated metrics designed to quantify \emph{improvement} (correctly added details not present at low resolution), \emph{omissions} (ground truth details missed by the SR method) and \emph{hallucinations} (incorrectly added, spurious details).

Surprisingly few works have assessed the usability of SR images for downstream analysis. A notable exception is the recent \cite{razzak2023multi}, where SR images are used for building delineation.
In this study, we use LC segmentation with a state-of-the-art foundation model as a representative task to evaluate the real-world usefulness of SR. The key idea is that SR is not an end in itself, but a tool to support image analysis. Its success should therefore be measured not by how realistic or visually appealing the images look, but by how well they can substitute true HR imagery in downstream analysis tasks. In doing so, we systematically demonstrate that some popular metrics are poor predictors of segmentation performance.

\section{Data}
\label{sec:data}

Our region of interest is the city of Hanoi, Vietnam — a dense, urban area. Urban areas often contain buildings and other small structures that are only tens of meters in size. Because of this, using free satellite data sources like Sentinel-2 or Landsat requires SR techniques. However, applying these techniques in urban settings is especially difficult.

We sidestep the use of synthetic data and train SR models on real, co-registered HR and LR images acquired with different satellite sensors. All images used were acquired between 2020 and 2023.

\subsection{Low-Resolution Imagery}
We use Sentinel-2 imagery \cite{esas2guide} as input,  in line with our objective to develop a SR method tailored to that sensor. Sentinel-2 data is freely available worldwide and provides consistent spatial and temporal coverage. It is collected by two identical satellites operating in a phase-shifted orbit, giving a revisit time of five days. This frequent coverage allows for multiple observations over a period of weeks or months, which supports MISR. We use the Level-2A surface reflectance product, accessed via SentinelHub \cite{sentinelhub}. To minimize differences due to land cover changed, we limit image acquisition to within two years of the corresponding HR target.

Of the 13 spectral bands observed by Sentinel-2, we only super-resolve the red, green, blue (RGB), and near infrared (NIR) bands, which are captured at 10$\,$m native resolution and overlaps with the spectral range of the HR Pl{\'e}iades Neo (PNEO) target. Reflectance values are clipped to the 2\textsuperscript{nd} and 98\textsuperscript{th} percentile to reduce the impact of cloud shadows and stray light effects \cite{Kuusk2016straylight}, then normalized to the $[0,1]$ range. Following \cite{worldstrat}, we divide the images into 373 square tiles, each covering an area of 2.5\,km\textsuperscript{2}, or 158\,$\times$\,158 pixels. Based on the findings from SATLAS~\cite{wolters2023zoomingzoominginadvancing}, we use eight input views for MISR. For each tile, the eight most suitable LR revisits are selected using three criteria: \emph{(i)} temporal proximity -— images taken closer to the date of the HR PNEO acquisition are preferred~\cite{Bai2016cloud}; \emph{(ii)} completeness --  images with lower cloud coverage and high quality pixels (non-defective, non-saturated, not covered by ice/snow) are favored, based on the provided cloud and scene classification masks~\cite{okabayashi2024LTAE}; and \emph{(iii)} spectral quality -- images with fewer pixels exceeding a reflectance value of 0.8 are preferred~\cite{Hannula2019snow}.

\begin{figure}[htbp]
    \includegraphics[width=\columnwidth]{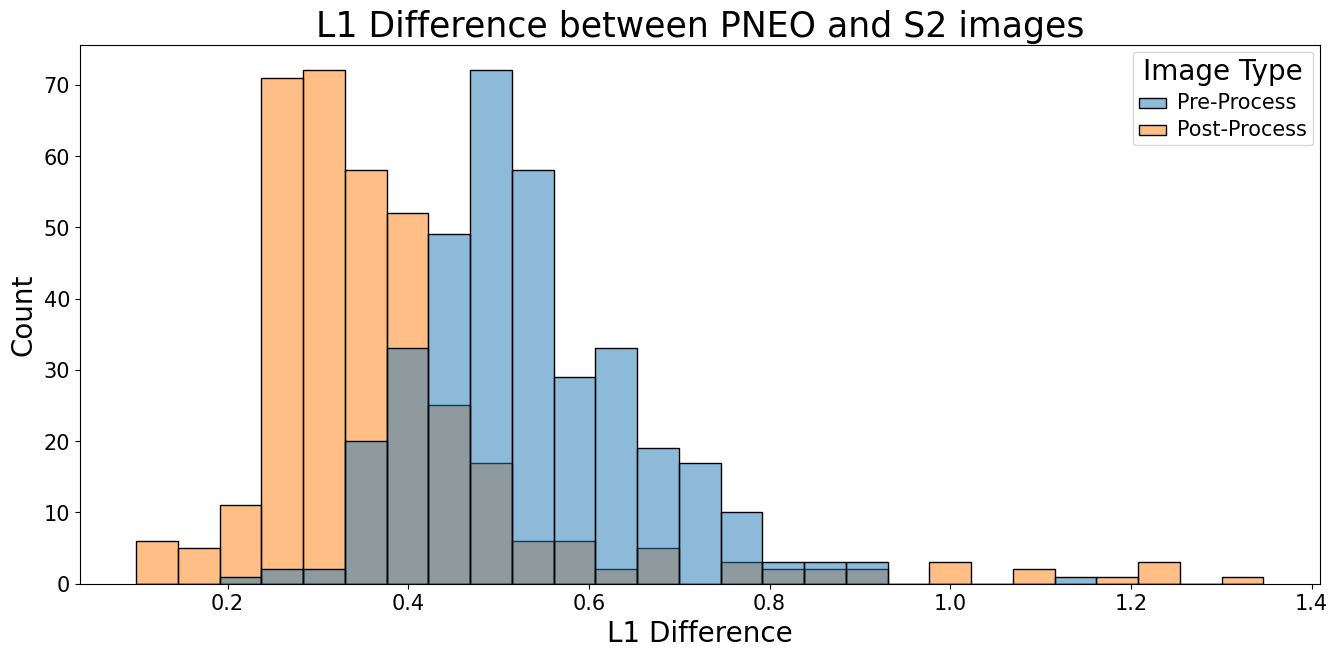}
    \caption{Histogram of $L_1$-differences between LR images and down-sampled HR images \textcolor{RoyalBlue}{before} and \textcolor{Peach}{after} radiometric cross-calibration (lower is better).}
    \label{fig:HISTOGRAM_RESULTS}
\end{figure}

\subsection{High-Resolution Imagery}
To ensure spectral consistency in the SR output, we use HR target images from a single sensor: the Pl\'eiades Neo (PNEO) constellation. PNEO was chosen over other HR sources due to its 1.2\,m native resolution, high spectral quality, and the availability of images with low off-nadir angles and minimal cloud cover over Hanoi. We use six top-of-atmosphere (TOA) HR images from Airbus OneAtlas~\cite{pneo}, which are divided into the same 2.5\,km\textsuperscript{2} square tiles as the Sentinel-2 data. Spectral values are normalized to the $[0, 1]$ range, then radiometrically aligned using histogram matching to the LR image from the same tile that best meets the three selection criteria described earlier. Finally, the tiles are bilinearly downsampled to a target resolution of 2.5\,m (632\,$\times$\,632 pixels).

\subsection{Land-Cover Labels}

Reference labels for LC classification were manually annotated on the basis of the HR images, using the QGIS software~\cite{QGIS_software}. Annotations are drawn as vector polygons, using the PNEO image tiles at native resolution to ensure the best visual quality. The class nomenclature comprises seven classes: buildings, sealed surfaces, water bodies, forest, grassland, cropland and bare soil.
The base tiles are center-cropped to 534$\times$534 pixels and overlaid with the Google Open Buildings dataset \cite{ggopenbuildings}, which was used to aid the annotation of buildings. 
Once completed, the labeled polygons are rasterized and downsampled to the 2.5$\,$m target resolution, retaining only those pixels whose neighboring pixels are uniformly covered by a single class, thus excluding mixed or ambiguous edge pixels.

\subsection{Training and Test Data Preparation}
For each of the eight selected LR images, masked pixels are imputed by averaging the valid reflectance values at the same location. The final RGB-NIR images are split into 64$\times$64 pixel patches with a sliding window and a stride of 48 pixels (25\% overlap between adjacent patches).

\begin{figure}[htbp]
    \includegraphics[width=\columnwidth]{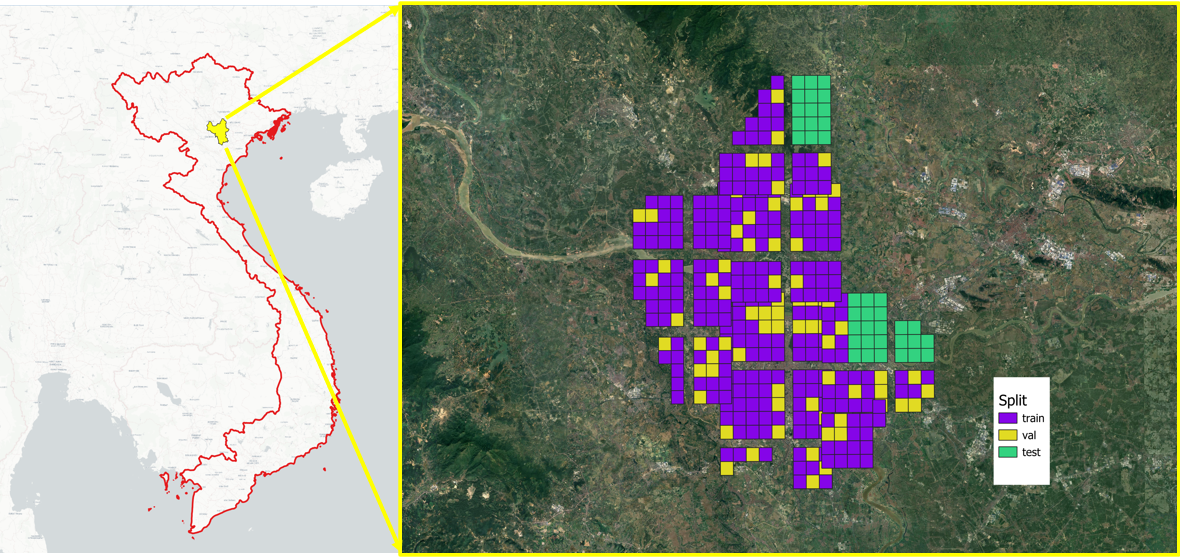}
    \caption{Training, validation and test regions of the Hanoi dataset.}
    \label{fig:SR_train_val_test}
\end{figure}

The resulting dataset is divided into training (70\%), validation (20\%), and test (10\%) portions, with geographic stratification to ensure a representative mix of urban, suburban and rural scenes. Two contiguous regions in the north and east are set aside as test data to minimize geographical correlation, see Figure~\ref{fig:SR_train_val_test}.

\section{Methods}

\subsection{\method{} Architecture}
Our SR network combines ideas from SISR and MISR in an integrated, end-to-end model. The single-image component encodes a latent prior distribution over high-resolution patterns from the training data and is responsible for reconstructing likely high-frequency details. Its design is inspired by Swin2SR~\cite{conde2022swin2sr}. The multi-image component integrates information from multiple satellite revisits, which are slightly shifted relative to one another due to small (unknown) geo-referencing errors. These shifts effectively oversample the surface reflectance and enable better reconstruction. This part of the network is based on the recursive fusion module of HighResNet~\cite{deudon2020highres}.

\begin{figure*}[htb]
\begin{subfigure}{0.533\textwidth}
\centering
\includegraphics[width=\linewidth]{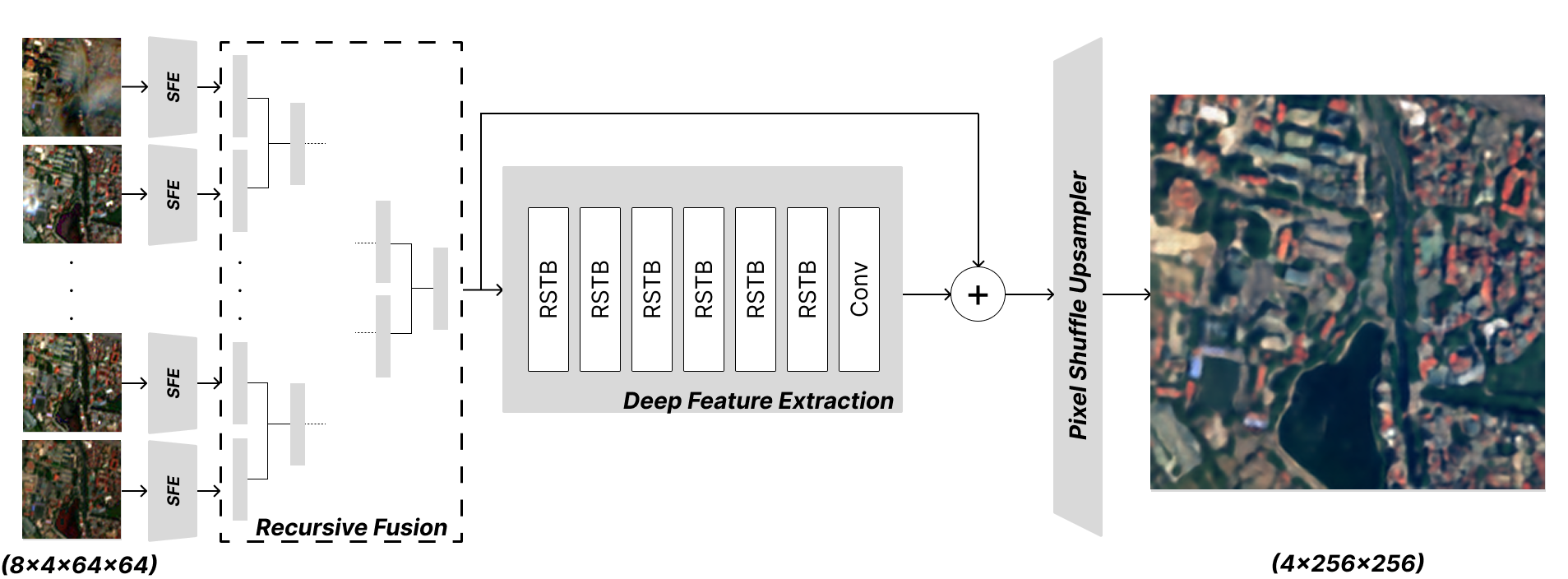}
\caption{\method{}}
\label{fig:PRE_DEEP_ARCH}
\end{subfigure}
\hfill
\begin{subfigure}{0.45\textwidth}
\centering
\raisebox{0.7mm}{
\includegraphics[width=\linewidth]{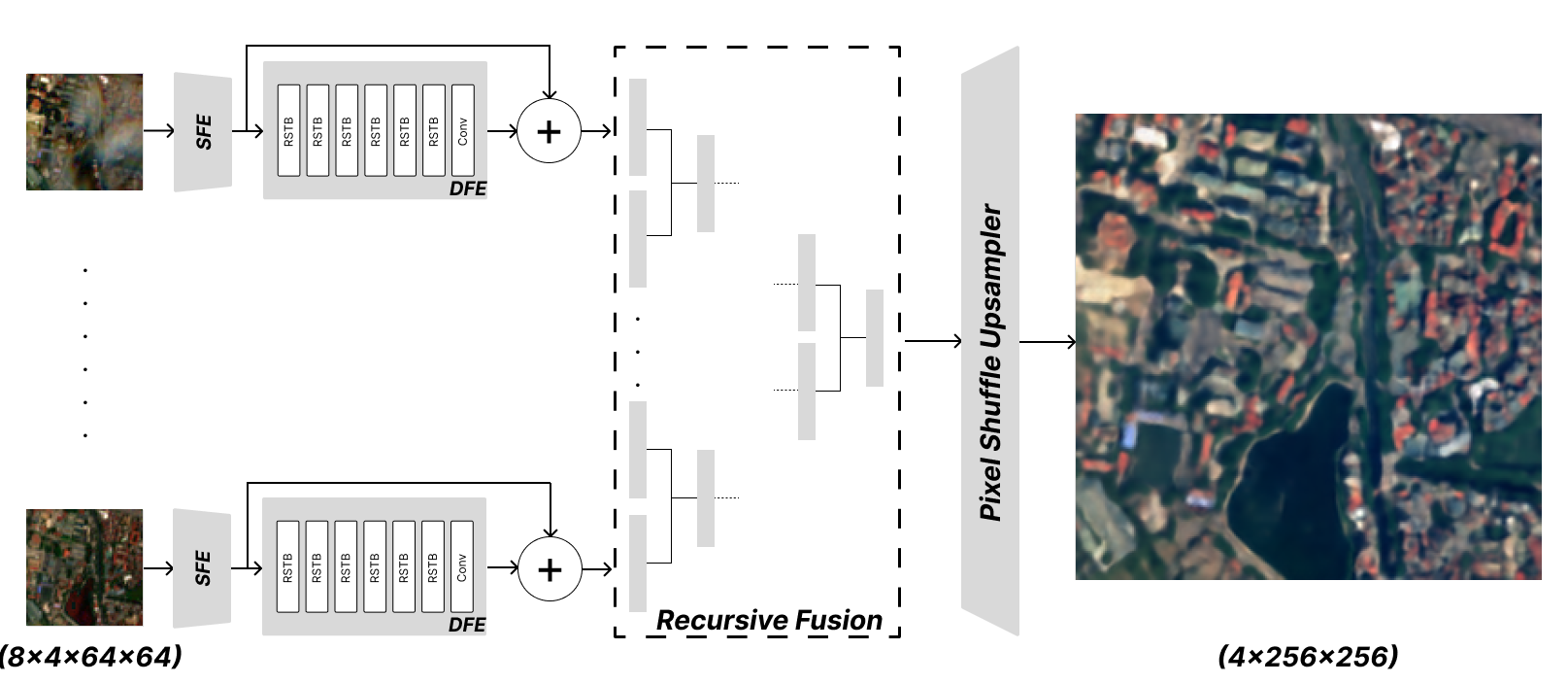}
}
\caption{\methodlate{}}
\label{fig:POST_DEEP_ARCH}
\end{subfigure}
 \caption{Architectures for combined SISR+MISR. Per default, we recommend the standard \method{}, where multi-image fusion precedes single-image enhancement of the fused feature representation.}   
\end{figure*}

The SISR component is a hybrid neural architecture consisting of two main stages: \emph{(i)} a \emph{deep feature extractor} with six residual Swin transformer blocks (RSTB) with windowed multi-head self-attention; and \emph{(ii)} 4$\times$ \emph{upsampling} through a pixel shuffle layer.
We follow the design of~\cite{conde2022swin2sr} and use six attention heads per layer, but modify the architecture by setting the window size to 8 and the embedding dimension to 258. Like the original Swin2SR, we include a 3$\times$3 convolutional \emph{shallow feature extractor} (SFE), which we place at the very beginning of the network—before the multi-image fusion step (see Figure~\ref{fig:PRE_DEEP_ARCH}).

To combine information from multiple satellite revisits, we use the recursive fusion strategy from HighResNet~\cite{deudon2020highres}. After the shallow feature extractor processes each input, we conduct pairwise merging whereby the feature maps from all eight LR views are combined until only one representation remains. Each pairwise merge applies the same fusion block: a two-layer convolutional residual block first updates both input feature maps, followed by a single residual convolution layer that merges them into one output feature map with the same number of channels.

While HighResNet makes effective use of the oversampling provided by multiple image acquisitions, its simple design lacks the representational capacity needed to model fine-grained high-resolution details. To address this, our experiments show that adding a high-capacity SISR module substantially improves performance. The complete \method{} model has approximately 30 million learnable parameters, of which $\approx$24 million bel ong to the SISR backbone.

We also evaluate an alternative design in which SISR is applied before MISR. In this variant, each input image is first processed independently using the Swin2SR backbone. The resulting feature maps are then recursively merged into a single representation, which is upsampled using a pixel shuffle layer (see Figure~\ref{fig:POST_DEEP_ARCH}). This late fusion approach, referred to as \methodlate{}, does not perform as well as the default early fusion strategy. As shown in Table~\ref{tab:performance_metrics}, it is also more computationally expensive, since the SISR backbone must be run separately for each input view.

\subsection{Implementation and Training Details}
All algorithms used in this work were implemented in PyTorch 2.4.0. For prior art we use the authors' original open-source implementations. We train \method{}, as well as all baseline models, from scratch on a single NVIDIA L4 GPU on Google Cloud Platform. The training configuration is consistent across all experiments: we use the Adam optimizer~\cite{adam2015} with an initial learning rate of $1 \cdot 10^{-4}$. Training is run for 100 epochs of 4 batches each, with a linear warm-up of the learning rate followed by a cosine annealing schedule.

\subsection{Land-Cover Model}
We use LC mapping as a representative spatial prediction task for evaluation. It is formulated as a standard pixel-wise semantic segmentation problem, where each pixel is assigned a LC class.
The segmentation network builds on the SATLAS foundation model \cite{bastani2023satlaspretrainlargescaledatasetremote}, which is a Swin-based encoder \cite{liu2022swintransformerv2scaling} pretrained on a large RGB dataset with resolutions between 0.5 and 2.5$\,$m. To accommodate our RGB-NIR data, we expand the model architecture to include a fourth channel and initialize the weights for the NIR channel as the average of the pretrained RGB weights. The model has 89,744,871 trainable parameters. 
The Swin encoder extracts feature maps at $\frac{1}{4}$, $\frac{1}{8}$, $\frac{1}{16}$, and $\frac{1}{32}$ of the 256$\times$256 pixel input patch size. The four feature maps are merged into a single latent representation with a feature pyramid network \cite{lin2017feature}. To extract segmentation maps, the representation is passed to a standard U-net decoder. 

The segmentation engine is trained on the same 2.5$\,$m PNEO images also used to train SR. For semantic segmentation we minimize a masked cross entropy loss, computed only at pixels with valid ground truth labels (i.e., excluding cloudy or unlabeled pixels). Training uses the Adam optimizer \cite{adam2015} with an initial learning rate $1 \cdot 10^{-4}$ that is dynamically adjusted using a cosine annealing scheduler, to a minimum of $1 \cdot 10^{-8}$. Training is conducted with a batch size of 16 for at most 1000 epochs, with early stopping when validation loss does not further decrease over 25 epochs.

\section{Experiments and Results}

\method{} is benchmarked against recent, high-performing super-resolution (SR) methods for Sentinel-2: Swin2SR~\cite{conde2022swin2sr} for single-image super-resolution (SISR); HighResNet~\cite{deudon2020highres} for multi-image super-resolution (MISR); and ESRGAN~\cite{wang2018esrgan}, a scheme~\cite{satlas} that simply stacks multiple images and feeds them into a popular SISR method. Note that the former two methods are closely related to the single- and multi-image components of \method{}, directly illustrating the impact of our combined scheme. As a trivial lower bound for SR and a sanity check, we also include bicubic upsampling of the low-resolution (LR) input.

For each SR method, we train on the training set (Section~\ref{sec:data}), apply the trained model to the images of the held-out test set, and perform all further evaluations on the high-resolution (HR) test images, at 2.5$\,$m resolution. To account for model uncertainty and the stochastic nature of neural network training, each method is trained five times using identical hyperparameters but different random seeds. We report the mean and empirical standard deviation of each evaluation metric across these five runs, providing a measure of both performance and uncertainty.

\subsection{Evaluation Metrics and Baseline Models}

We evaluate the quality of the SR images (at 2.5\,m resolution) by applying a land-cover (LC) classifier and measuring the accuracy of the resulting classification maps. Accuracy is assessed using two standard segmentation metrics: overall accuracy (the fraction of correctly classified pixels) and the mean intersection-over-union (mIoU), which balances performance across all classes. Unlike overall accuracy, mIoU is not dominated by the more frequent classes.

There are two common definitions of IoU: the \emph{macro} IoU, which averages the per-class IoU scores, and the \emph{micro} IoU, which computes the IoU over all pixels without class distinction. We use the macro IoU as our primary metric because it better reflects performance on underrepresented classes, though we also report micro IoU for completeness.

To provide an upper bound on achievable performance, we run the same LC classifier on the harmonized PNEO images at 2.5\,m resolution. When replacing true HR images with SR ones, a drop in performance is expected. The size of this gap reflects how well SR methods can close the resolution gap between free LR data and expensive HR imagery in the context of land-cover mapping.

In addition to downstream task performance, we also compute conventional image quality metrics: PSNR, SSIM, LPIPS, and the hallucination, improvement, and omission metrics from OpenSRTest~\cite{aybar2024opensr}. However, as our results show, higher scores on these image-level metrics do not consistently correlate with better LC classification accuracy.

\begin{table*}[!tb]
\caption{Results for high-resolution land-cover classification.}
\vspace{0.5em}
\label{tab:sr_lulc_metrics}
\centering
\begin{tabularx}{\textwidth}{|X|X|>{\centering\arraybackslash}X>{\centering\arraybackslash}X>{\centering\arraybackslash}X|}
\hline
\textbf{SR Method} &
\textbf{Type} &
\textit{Acc\,{\tiny$\uparrow$}} &
\textit{mIoU\,{\tiny$\uparrow$}} & \textit{mIoU\textsubscript{micro}\,{\tiny$\uparrow$}} \\
\hline
\textcolor{gray}{HR Image} &
\textcolor{gray}{upper bound} & 
\textcolor{gray}{0.856}
\scriptsize{\textcolor{white}{ $\pm$  0.000}} &
\textcolor{gray}{0.663}
\scriptsize{\textcolor{white}{ $\pm$  0.000}} & \textcolor{gray}{0.748}
\scriptsize{\textcolor{white}{ $\pm$  0.000}} \\
\textcolor{gray}{Majority Class} &
\textcolor{gray}{learning-free baseline} &
\textcolor{gray}{0.313}
\scriptsize{\textcolor{white}{ $\pm$  0.000}} & \textcolor{gray}{0.045}
\scriptsize{\textcolor{white}{ $\pm$  0.000}} & \textcolor{gray}{0.185}
\scriptsize{\textcolor{white}{ $\pm$  0.000}} \\
\textcolor{gray}{Bicubic} &
\textcolor{gray}{learning-free baseline} &
\textcolor{gray}{0.440}
\scriptsize{\textcolor{white}{ $\pm$  0.000}} & \textcolor{gray}{0.278}
\scriptsize{\textcolor{white}{ $\pm$  0.000}} & \textcolor{gray}{0.282}
\scriptsize{\textcolor{white}{ $\pm$  0.000}} \\
Swin2SR & SISR &
0.714\scriptsize{ $\pm$ 0.002} &
0.489\scriptsize{ $\pm$ 0.003} &
0.555\scriptsize{ $\pm$ 0.002}\\
HighResNet & MISR &
0.583\scriptsize{ $\pm$ 0.005} &
0.387\scriptsize{ $\pm$ 0.004} &
0.411\scriptsize{ $\pm$ 0.005} \\
ESRGAN & hybrid &
0.724\scriptsize{ $\pm$ 0.010} &
0.493\scriptsize{ $\pm$ 0.010} &
0.567\scriptsize{ $\pm$ 0.012}\\
\method{} & hybrid &
\textbf{0.746}\scriptsize{ $\pm$ 0.001} & \textbf{0.516}\scriptsize{ $\pm$ 0.003} & \textbf{0.595}\scriptsize{ $\pm$ 0.002} \\
\methodlate{} & hybrid & 
0.728\scriptsize{ $\pm$ 0.001} &
0.500\scriptsize{ $\pm$ 0.002} &
0.572\scriptsize{ $\pm$ 0.002} \\
\hline
\end{tabularx}
\end{table*}

\subsection{LC Classification Results}

The accuracies of LC classification with different inputs are reported in Table \ref{tab:sr_lulc_metrics}. First of all, they confirm that -- unsurprisingly -- higher image resolution benefits the mapping task: the segmentation of PNEO reaches 85.6\% accuracy, respectively 66.3\% mean IoU on the test set. In contrast, segmenting the bicubically upsampled Sentinel-2 image yields 44.0\% accuracy and 27.8\% mIoU. In other words, the classifier largely fails on naïvely upsampled images with very different local contrast statistics (especially since a naïve solution where every pixel is labeled as cropland reaches 31.3\% accuracy).

As expected, all SR models outperform the bicubic baseline, but none match the performance achievable with real PNEO images. This supports the claim that SR can partially reduce the domain gap between LR inputs and HR targets.

Among the tested SR methods, \method{} achieves the highest performance. It improves segmentation accuracy for most classes over the other methods (see Table~\ref{tab:perclass}), reaching an overall accuracy of 74.6\% and an average mIoU of 51.6\%.

Somewhat unexpectedly, the single-image Swin2SR outperforms the multi-image HighResNet. This suggests that the blur introduced by multi-image fusion may hinder segmentation performance more than the artificial high-frequency details generated by perceptual and adversarial losses. Hybrid methods that combine multi-image fusion with sufficient model capacity to encode a strong image prior perform best. Among these, \method{} leads by a sinificant margin of 2.3 percentage points in mIoU over ESRGAN.

The performance gap between \method{} and Swin2SR is likely due to the inclusion of the MISR component, supporting the advantage of multi-image fusion in SR. The smaller gap between \method{} and ESRGAN may have several causes. One possibility is that simple input stacking is less effective than a dedicated, recursive fusion mechanism. Another is that ESRGAN’s architecture may not fully leverage the training data, either due to limited capacity or the notorious instability of adversarial training.

We also observe that our late fusion variant \methodlate{} performs on par with recent  SR methods, but does not exceed them. This could be due to two factors: first, late fusion can introduce blur at a point where no further processing layers are available to correct it; second, applying SR independently to each input may produce inconsistent high-frequency details, which are difficult to reconcile during fusion.

\begin{table*}[tb]
\caption{Per-class accuracies of land-cover classification.}
\vspace{-0.5em}
\label{tab:perclass}
\begin{center}
\resizebox{\textwidth}{!}{%
\begin{tabular}{|l|l|ccccccc|}
\hline
\textbf{SR Method} &
\textbf{Type} &
\textit{Buildings} &
\textit{Sealed} &
\textit{Water} & 
\textit{Forest} &
\textit{Grassland} &
\textit{Crop}&
\textit{Bare Soil}\\
\hline
\textcolor{gray}{HR Image} &
\textcolor{gray}{upper bound} &
\textcolor{gray}{0.910}
\scriptsize{\textcolor{white}{ $\pm$  0.000}} & \textcolor{gray}{0.707}
\scriptsize{\textcolor{white}{ $\pm$  0.000}} &
\textcolor{gray}{0.898}
\scriptsize{\textcolor{white}{ $\pm$  0.000}} & \textcolor{gray}{0.879}
\scriptsize{\textcolor{white}{ $\pm$  0.000}} & \textcolor{gray}{0.576}
\scriptsize{\textcolor{white}{ $\pm$  0.000}} & \textcolor{gray}{0.929}
\scriptsize{\textcolor{white}{ $\pm$  0.000}}
& \textcolor{gray}{0.542}
\scriptsize{\textcolor{white}{ $\pm$  0.000}}
\\
\textcolor{gray}{Majority Class} &
\textcolor{gray}{baseline} & 
\textcolor{gray}{0.000}
\scriptsize{\textcolor{white}{ $\pm$  0.000}} & \textcolor{gray}{0.000}
\scriptsize{\textcolor{white}{ $\pm$  0.000}} &
\textcolor{gray}{0.000}
\scriptsize{\textcolor{white}{ $\pm$  0.000}} & \textcolor{gray}{0.000}
\scriptsize{\textcolor{white}{ $\pm$  0.000}} & \textcolor{gray}{0.000}
\scriptsize{\textcolor{white}{ $\pm$  0.000}} & \textcolor{gray}{1.000}
\scriptsize{\textcolor{white}{ $\pm$  0.000}}
& \textcolor{gray}{0.000}
\scriptsize{\textcolor{white}{ $\pm$  0.000}}\\
\textcolor{gray}{Bicubic} &
\textcolor{gray}{baseline} & 
\textcolor{gray}{0.243}
\scriptsize{\textcolor{white}{ $\pm$  0.000}} & \textcolor{gray}{0.212}
\scriptsize{\textcolor{white}{ $\pm$  0.000}} &
\textcolor{gray}{0.384}
\scriptsize{\textcolor{white}{ $\pm$  0.000}} & \textcolor{gray}{0.594}
\scriptsize{\textcolor{white}{ $\pm$  0.000}} & \textcolor{gray}{0.764}
\scriptsize{\textcolor{white}{ $\pm$  0.000}} & \textcolor{gray}{0.488}
\scriptsize{\textcolor{white}{ $\pm$  0.000}} 
& \textcolor{gray}{0.548}
\scriptsize{\textcolor{white}{ $\pm$  0.000}}\\

Swin2SR & SISR &
0.667\scriptsize{ $\pm$ 0.006} &
0.434\scriptsize{ $\pm$ 0.002} &
0.848\scriptsize{ $\pm$ 0.002} &
0.699\scriptsize{ $\pm$ 0.006} &
0.472\scriptsize{ $\pm$ 0.037} &
0.828\scriptsize{ $\pm$ 0.003} &
0.527\scriptsize{ $\pm$ 0.004} \\
HighResNet & MISR &
0.438\scriptsize{ $\pm$ 0.003} &
0.332\scriptsize{ $\pm$ 0.003} &
0.759\scriptsize{ $\pm$ 0.003} &
0.589\scriptsize{ $\pm$ 0.020} &
\textbf{0.615}\scriptsize{ $\pm$ 0.026} &
0.629\scriptsize{ $\pm$ 0.005} &
\textbf{0.611}\scriptsize{ $\pm$ 0.008} \\
ESRGAN & hybrid &
0.705\scriptsize{ $\pm$ 0.020} &
0.422\scriptsize{ $\pm$ 0.011} &
0.803\scriptsize{ $\pm$ 0.002} &
0.755\scriptsize{ $\pm$ 0.003} &
0.361\scriptsize{ $\pm$ 0.005} &
0.869\scriptsize{ $\pm$ 0.017} &
0.450\scriptsize{ $\pm$ 0.006} \\
\method{} & hybrid &
0.688\scriptsize{ $\pm$ 0.009} & \textbf{0.483}\scriptsize{ $\pm$ 0.002} &
\textbf{0.853}\scriptsize{ $\pm$ 0.004} &
\textbf{0.763}\scriptsize{ $\pm$ 0.000} &
0.343\scriptsize{ $\pm$ 0.022}&
\textbf{0.897}\scriptsize{ $\pm$ 0.007}&
0.508\scriptsize{ $\pm$ 0.011}\\
\methodlate{} & hybrid &
\textbf{0.708}\scriptsize{ $\pm$ 0.006} &
0.454\scriptsize{ $\pm$ 0.002} &
0.850\scriptsize{ $\pm$ 0.001} &
0.697\scriptsize{ $\pm$ 0.007} &
0.417\scriptsize{ $\pm$ 0.013} & 
0.856\scriptsize{ $\pm$ 0.003} & 
0.511\scriptsize{ $\pm$ 0.007}\\
\hline
\end{tabular}%
}
\end{center}
\end{table*}

\begin{table*}[tb]
\caption{Quantitative results in terms of image quality and super-resolution metrics.}
\vspace{0.5em}
\label{tab:combined_sr_metrics}
\centering
\begin{tabularx}{\textwidth}{|l|l|>{\centering\arraybackslash}X>{\centering\arraybackslash}X>{\centering\arraybackslash}X|>{\centering\arraybackslash}X>{\centering\arraybackslash}X>{\centering\arraybackslash}X|}
\hline
\textbf{SR Method} & \textbf{Type} &
\textit{PSNR {\tiny$\uparrow$}} &
\textit{SSIM {\tiny$\uparrow$}} &
\textit{LPIPS {\tiny$\downarrow$}} & 
\textit{Ha. {\tiny$\downarrow$}} &
\textit{Om. {\tiny$\downarrow$}} &
\textit{Im. {\tiny$\uparrow$}} \\
\hline
\textcolor{gray}{Bicubic} &
\textcolor{gray}{baseline} &
\mbox{\textcolor{gray}{16.006}
\scriptsize{\textcolor{white}{ $\pm$  0.000}}} & \textcolor{gray}{0.369}
\scriptsize{\textcolor{white}{ $\pm$  0.000}} &
\textcolor{gray}{0.547}
\scriptsize{\textcolor{white}{ $\pm$  0.000}} & \textcolor{gray}{0.236}
\scriptsize{\textcolor{white}{ $\pm$  0.000}} & \textcolor{gray}{0.572}
\scriptsize{\textcolor{white}{ $\pm$  0.000}} & \textcolor{gray}{0.193}
\scriptsize{\textcolor{white}{ $\pm$  0.000}}\\
Swin2SR & SISR &
\mbox{16.397\scriptsize{ $\pm$ 0.009}} &
0.411\scriptsize{ $\pm$ 0.001} &
0.470\scriptsize{ $\pm$ 0.000} &
0.299\scriptsize{ $\pm$ 0.001} &
0.377\scriptsize{ $\pm$ 0.002} &
0.324\scriptsize{ $\pm$ 0.002}\\
HighResNet & MISR &
\mbox{\textbf{16.968}\scriptsize{ $\pm$ 0.004}} &
0.415\scriptsize{ $\pm$ 0.001} &
0.490\scriptsize{ $\pm$ 0.001} &
\textbf{0.270}\scriptsize{ $\pm$ 0.001} &
0.388\scriptsize{ $\pm$ 0.006} &
0.342\scriptsize{ $\pm$ 0.005}\\
ESRGAN & hybrid &
\mbox{16.630\scriptsize{ $\pm$ 0.022}} &
0.416\scriptsize{ $\pm$ 0.002} &
0.459\scriptsize{ $\pm$ 0.002} &
0.285\scriptsize{ $\pm$ 0.001} &
0.339\scriptsize{ $\pm$ 0.004} &
0.376\scriptsize{ $\pm$ 0.005}\\
\method{} & hybrid &
\mbox{16.676\scriptsize{ $\pm$ 0.017}} &
\textbf{0.419}\scriptsize{ $\pm$ 0.002} & \textbf{0.444}\scriptsize{ $\pm$ 0.000} &
0.286\scriptsize{ $\pm$ 0.001} &
\textbf{0.331}\scriptsize{ $\pm$ 0.004} & \textbf{0.383}\scriptsize{ $\pm$ 0.004}\\
\methodlate{} & hybrid &
\mbox{16.468\scriptsize{ $\pm$ 0.010}} &
0.416\scriptsize{ $\pm$ 0.001} &
0.461\scriptsize{ $\pm$ 0.000} &
0.294\scriptsize{ $\pm$ 0.001} &
0.373\scriptsize{ $\pm$ 0.002} &
0.333\scriptsize{ $\pm$ 0.002}\\
\hline
\end{tabularx}
\end{table*}

\subsection{Image Quality Metrics}

For all evaluated methods, we compute both standard image quality metrics with respect to the PNEO ground truth and specialized SR metrics provided by the OpenSR-test benchmark. As shown in Table~\ref{tab:combined_sr_metrics}, all SR models outperform naïve bicubic upsampling across nearly all metrics, except for hallucination scores, which are naturally lowest when no high-frequency content is introduced at all. A key observation is that the ranking of SR methods varies depending on the chosen metric.

Metrics such as the widely used PSNR and the hallucination score are particularly poor indicators of downstream utility. For example, HighResNet achieves strong results on these metrics despite weaker segmentation performance. This is likely because such metrics favor smooth, blurrier images, which avoid penalties for small misalignments or high-frequency artifacts.

SSIM generally has limited discriminative power. All methods except bicubic upsampling achieve nearly identical scores, with only a small drop observed for Swin2SR.

LPIPS is the metric that best reflects the relative segmentation performance. Only \methodlate{} and ESRGAN are out of order; however, the difference in LPIPS scores between them is hardly significant. We hypothesize that LPIPS matches segmentation performance best because it measures perceptual similarity in the feature space of a convolutional neural network, which may be more aligned with the features used by the segmentation model.

Turning to the OpenSR metrics, the \textit{hallucination score} is highest for Swin2SR, which represents pure single-image SR, and lowest for HighResNet, which uses multi-image fusion, as expected. The \textit{omission score} largely mirrors the segmentation performance ranking, with only \methodlate{} out of order. The \textit{improvement score}, on the other hand, gives an overly optimistic assessment of HighResNet's performance and fails to capture the substantial gap between ESRGAN and \method{}.

Overall, our findings suggest that most generic image quality metrics, including PSNR, SSIM, and even the custom-designed SR improvement score, are not reliable indicators of how well SR images support downstream tasks such as semantic segmentation. LPIPS is the only tested metric that correctly reflects suitability for land-cover mapping.

While further research is needed to determine whether this limitation also applies to other image analysis tasks, our results raise concerns about the common practice of evaluating and tuning SR models based solely on image quality metrics.

\subsection{Qualitative Evaluation}
\begin{figure*}[htbp]
    \centering
\includegraphics[width=.93\textwidth, keepaspectratio]{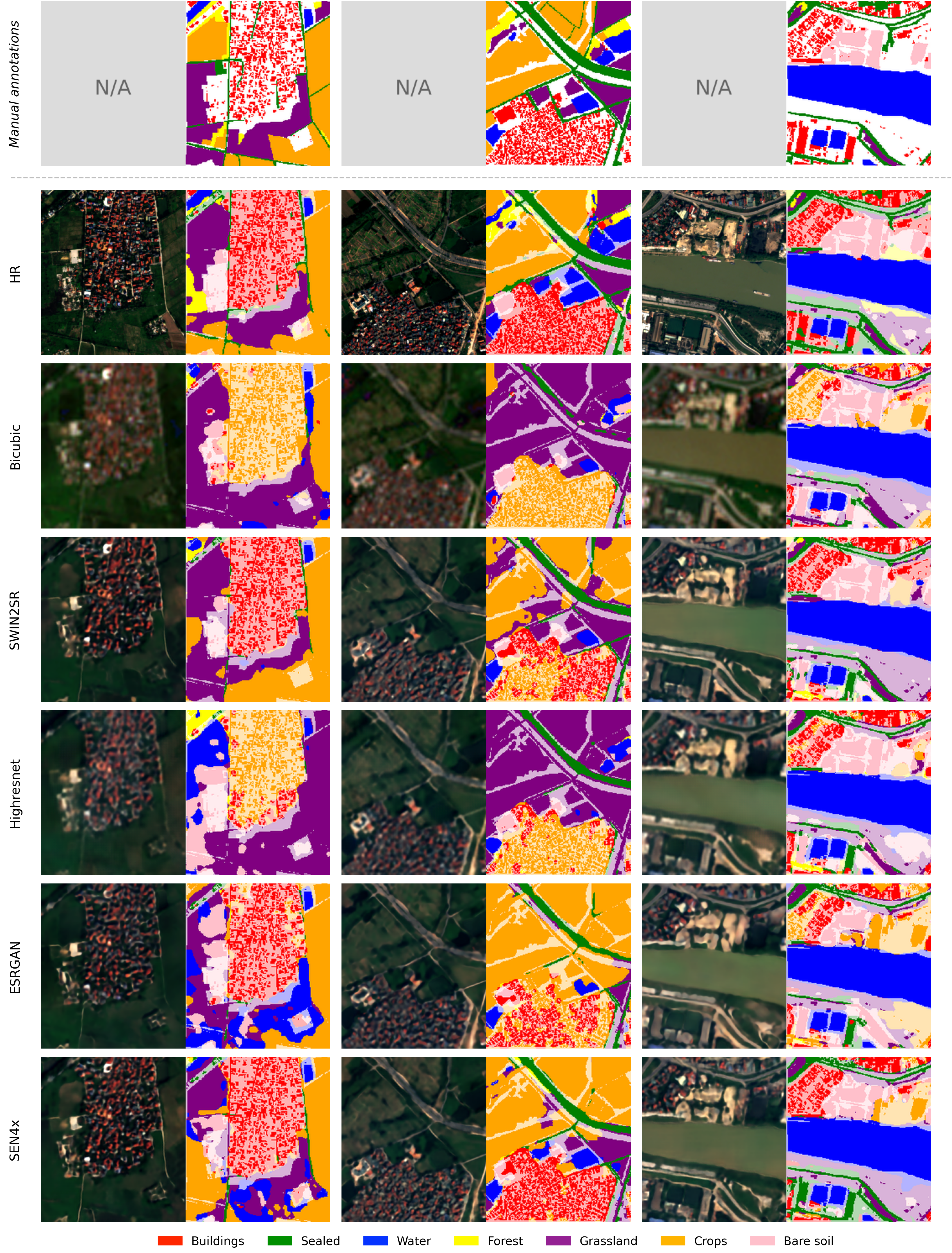}
    \caption{Comparison of SR results and LC segmentations for three exemplary scenes from Hanoi. Regions without ground truth labels are denoted by transparent masks.}
    \label{fig:SR_IMAGE_RESULTS}
\end{figure*}

Figure~\ref{fig:SR_IMAGE_RESULTS} provides a side-by-side comparison of SR images (only RGB channels) and their corresponding LC maps, for three selected scenes. Rows correspond to different SR methods.

\begin{table}[!b]
\caption{Wall Clock Time for super-resolving one 64$\times$64 pixel Sentinel-2 patch on Google Cloud.}
\vspace{0.5em}
\label{tab:performance_metrics}
\centering
\begin{tabularx}{\columnwidth}{|l|l|>{\centering\arraybackslash}Xc|}
\hline
\textbf{SR Method} & \textbf{Type} & \mbox{\textit{\# Trainable Params}} & \textit{Time (ms)} \\
\hline
Swin2SR & SISR & 24,525,082 & \textcolor{white}{0}133.6\scriptsize{ $\pm$\textcolor{white}{.0}7.9}  \\
HighResNet & MISR & 12,991,084 & \textcolor{white}{00}65.1\scriptsize{ $\pm$\textcolor{white}{.0}4.8}\\
ESRGAN & hybrid & 16,715,268 & \textcolor{white}{00}31.9\scriptsize{ $\pm$\textcolor{white}{.0}3.6} \\
\method{} & hybrid & 30,517,135 & \textcolor{white}{0}189.6\scriptsize{ $\pm$\textcolor{white}{.0}2.6} \\
\methodlate{} & hybrid & 30,517,135 & 1471.5\scriptsize{ $\pm$\textcolor{white}{.}30.2} 
\\ 
\hline
\end{tabularx}
\end{table}

Across all three examples, learned models achieve a clear improvement over the bicubic baseline. Looking at the two best-performing models, \method{} and ESRGAN, the visual differences are small. Nevertheless, the LC maps derived from them are noticeably different -- particularly in regions characterized by high-frequency details. 
In the first column, \method{} more reliably distinguishes vegetation from water bodies, likely due to a more faithful reconstruction of color. Note also the visibly better recovery of thin features such as roads. In the center and second example scene, \method{} handles small, densely spaced buildings more accurately.
The visualizations also highlight an important advantage of SR that is not fully captured by global performance metrics. Segmentation of large, homogeneous regions, such as grasslands or water bodies, is often adequate even in low-resolution inputs or with basic SR methods. In contrast, segmentation errors are more common on small structures, such as roads or buildings. These minority classes, however, are often of particular importance in urban remote sensing applications.

\subsection{Computational Cost}
In Table ~\ref{tab:performance_metrics} we report the number of trainable parameters for all models, as well as computation times. For the latter, we show the average and standard deviation of the wall clock time needed to perform inference for one $4\times64\times64$ patch on a Google Cloud virtual machine with 8vCPU, 30 GB RAM, and one NVIDIA T4 GPU.

\section{Conclusion}

We have studied the integration of single-image (SISR) and multi-image super-resolution (MISR) techniques to enhance the spatial resolution of Sentinel-2 imagery, and the potential of super-resolved (SR) imagery for an elementary image interpretation task, semantic segmentation. Our hybrid \method{} architecture effectively combines the multi-image data fusion of MISR with the strong image prior of SISR, leading to significant improvements of a subsequent land-cover (LC) classification. Our study confirms that SR imagery can narrow the performance gap between freely available Sentinel-2 data and costly high-resolution (HR) imagery, with implications for scalable and cost-efficient geospatial analysis.

In addition, while standard image quality metrics such as PSNR and SSIM remain widely used for evaluating SR outputs, our results show that they correlate poorly with the usefulness of SR images for downstream analysis. In many cases, they fail to reflect relevant differences between models. These findings support the adoption of task-specific performance metrics as a more appropriate measure of SR quality.

Several limitations should be noted. First, the study is geographically restricted to Hanoi, Vietnam. Future work should investigate generalization across wider geographic areas. Second, our SR models were trained and evaluated using only four Sentinel-2 bands (RGB + NIR). While these bands are commonly used and sufficient for many tasks, excluding additional spectral bands (such as red-edge or short wave infrared) risks losing important information. Extending SR to those bands could enhance the approach, but remains challenging due to the lack of high-resolution reference data. Third, we have evaluated SR through a single downstream task, LC classification. Although this is a fundamental task in Earth observation, future studies should explore the potential of SR for other applications.

In conclusion, our work highlights the advantages of a hybrid SISR and MISR approach, as well as their practical benefits in the context of open remote sensing images. Analysis-ready SR images can bring substantial improvements for subsequent analysis and, in some cases, serve as a viable alternative to HR data.

Finally, we advocate for a shift in SR evaluation practices. We argue that evaluation should focus on task-specific utility.
Rather than relying on reconstruction errors or subjective visual quality, the effectiveness of SR is best judged by its contribution to the downstream information extraction that motivates satellite image analysis in the first place.

\section*{Acknowledgment}
The grant fund for this work was received from the Japan Fund for
Prosperous and Resilient Asia and the Pacific financed by the Government of
Japan through the Asian Development Bank. The European Space Agency's Global Development Assistance programme provided in-kind support for GeoVille's work on image labelling and land use classification. The authors especially thank Hanna Koloszyc and Julieta Bolgeri of GeoVille for their support. Additionally, the Pl\'eiades Neo images used in the analysis were provided by the European Space Agency's Third Party Missions programme. The authors also thank Julia Roque for her assistance. The views and conclusions presented in this paper are those of the authors and do not necessarily reflect the official policies or positions of the Asian Development Bank, its Board of Governors, or the governments they represent.

\section*{Code and Data Availability}
All code, trained models and data preparation scripts used in this study will be made publicly available at:
\url{https://github.com/ADB-Data-Division/sen4x}.

\clearpage
\bibliographystyle{elsarticle-num}
\bibliography{adbsr.bib}

\end{document}